\documentstyle[aps,prl,multicol]{revtex}
\input{epsf.tex}
\begin{document}
\draft
\title{Comment on "Collective excitations of a degenerate gas at the 
BEC-BCS crossover"}
\author{R. Combescot and X. Leyronas}
\address{Laboratoire de Physique Statistique,
 Ecole Normale Sup\'erieure*,
24 rue Lhomond, 75231 Paris Cedex 05, France}
\date{Received \today}
\maketitle


\begin{multicols}{2}
Very recent experiments have studied for the first time collective 
excitations of an ultracold $^6$Li gas \cite{thomas,grim}, covering in 
particular the BEC-BCS crossover domain \cite{grim}. We wish to 
point out that the results for the axial mode, through hydrodynamics, 
give direct access to the (3D) equation of state of the strongly interacting 
gas, mostly near the unitarity limit. On the other hand the surprising 
results found for the radial mode are actually not necessarily in 
contradiction with the expectations from superfluid hydrodynamics.

Indeed the radial mode frequencies $ \Omega _r$ are near \cite{thomas} 
$2 \pi . 2400$ Hz and \cite{grim}  $2 \pi . 1200$ Hz. The frequency $ 
\omega _F = E_F/ \hbar$ corresponding to the Fermi energy is 
\cite{grim} $1.6$ $10 ^{5}$ Hz. If on the BCS side we estimate the 
gap by the BCS relation $ \Delta = 1.75 T_c $ and take a nearly optimal 
value $ T_c = 0.2 E_F$, we have $ \hbar \Omega _r  / \Delta \sim 0.13 
$ at the trap center, this ratio increasing when we go away from the 
center. This is large enough to make questionable hydrodynamics, 
which assumes  $\hbar  \Omega_r  / \Delta \ll 1$ in order to be accurate, 
and may explain the 10\% discrepancy between theory and experiment 
at unitarity. Similarly one can estimate that the ratio $ \xi / l_{\perp}$  
between the Cooper pair size and the transverse size of the trap is at best 
given by this same figure $\sim 0.1 $, which makes doubtful the 
accuracy of the local density approximation (a necessary ingredient in 
the hydrodynamic result). The situation in Ref. \cite{thomas} is even 
worse. This is also consistent with a natural superfluid interpretation of 
the strong attenuation \cite{grim} at $910$ G as a pair-breaking peak 
corresponding to $ \hbar \Omega _r = 2 \Delta (T,B)$, occuring 
because $ T_c$ and $ \Delta $ decrease with increasing field $B$. 
Clearly in this case $ \hbar \Omega _r  / \Delta $ is no longer small. 
Taking for example $ \Delta (T,B) \simeq T_c$ (the gap depends on 
$T/T_c$, which is not known) gives $T_c/E_F \simeq 0.02$, coherent 
with the estimated temperature $T$ in this experiment.

On the other hand the axial mode is a very good case for making use of 
the hydrodynamic limit. Indeed its frequency is very low $\hbar  
\Omega_a  / \Delta \sim 5. 10 ^{-3}$, which is the appropriate range for 
this approximation. Moreover the experimental temperature is certainly 
quite low, as confirmed by the very low damping found in most of the 
magnetic field range. This makes it possible to neglect the effect of 
dissipation on the frequency of the modes. Finally for the very 
elongated traps used in experiments, one has to deal with a simple 
effective one-dimensional problem. In this case, when the chemical 
potential $ \mu (n)$ of the (3D) gas is a power law of the total particle 
density $ n ^{1/p}$, the axial mode frequency is given by an exact 
analytical result \cite{str1,ms} $ \omega ^{2}/ \omega ^{2}_{z}= 2 + 
1/(p+1) $. In the present situation this case is found in the BEC limit 
(small positive scattering length $a$), with $ p = 1$, leading to $ 
\omega ^{2}/ \omega ^{2}_{z}= 5/2 $, in the BCS limit (small 
negative scattering length $a$), with $ p = 3/2 $ and $ \omega ^{2}/ 
\omega ^{2}_{z}= 12/5 $. Moreover quite remarkably \cite{str2} this 
same value  $ p = 3/2$ applies also in the unitarity limit, where $a$ is 
very large. These values are in fair agreement with experiment 
\cite{grim} for the BEC and the unitarity cases.  

We have shown recently \cite{rcxl} how it is possible to invert such 
experimental data to obtain the equation of state of the gas. This 
requires only that $ \mu (n)$ is known in some limiting case, from 
which one then go away by an iterative procedure, making use of the 
experimental knowledge of the mode frequency as a function of 
density. The basic ingredient of this method has been shown recently 
\cite{fuchs} to have an accuracy of order $ 10 ^{-3}$. In the present 
case we have in principle the choice between the three limiting cases 
mentionned above, the most convenient one being the unitarity limit. 
However there are not enough data to carry out the above program with 
a sensible precision. Nevertheless we can analyze the region in the 
vicinity of the unitarity limit, where a fairly linear behaviour is obtained 
experimentally. In this case we have the simpler problem of performing 
a perturbative calculation \cite{fuchs}. 
\vspace{-5mm}
\begin{figure}
\centering
\vbox to 55mm{\hspace{-6mm} \epsfysize=55mm
\epsfbox{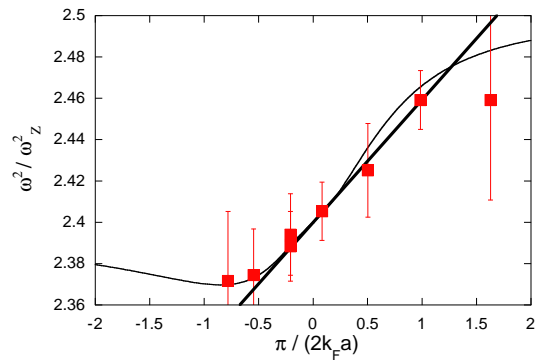} }
\caption{Reduced axial mode frequency as a function
of the inverse scattering length $a^{-1}$ for the model
in the text. Heavy line: linear approximation near unitarity. Red squares: 
experimental data of Ref. 2.}
\label{figure1}
\end{figure}
\vspace{-2mm}
In the vicinity of the unitarity limit we linearize the general form $ \mu 
(n) = \hbar ^{2} k_F ^{2} f(1/k_F a) /2m$ of the chemical potential by 
$f(y)= \xi -Sy$, where $ k_F ^{3}=3 \pi ^{2}n $ and $ \xi $ and $S$ 
are constants. Using the method of Ref. \cite{fuchs}, we find \cite{bulg} 
that, in 
this regime, the shift of the mode frequency is given analytically by 
$\delta ( \omega ^{2}/ \omega ^{2}_{z}) \equiv \omega ^{2}/ \omega 
^{2}_{z} - 12/5 = (256/875 \pi ) (S/ \xi ) ( 1/ k _{Fmax} a) $ where $ 
k _{Fmax}$ is the (3D) Fermi wavevector at the center of the trap. 
Taking \cite{grim} $E_F=1.2 \mu K$ and $\delta  (1/a) = 5.7$ $ 10 
^{4}m ^{-1}G ^{-1}$ in the vicinity of the resonance, this gives  
$\delta ( \omega ^{2}/ \omega ^{2}_{z}) \simeq 10 ^{-3} S/ \xi G ^{-
1}$. Comparing with the experimental result $ \simeq 1.1$ $ 10 ^{-3} 
G ^{-1}$ we obtain $S/ \xi = 1.1$. If we take for $ \xi \equiv 1+ \beta 
$ a value \cite{str2} $ \xi = 0.45 $ which is most likely both 
experimentally \cite{bourdel} and theoretically \cite{carl}, we find the 
experimental result $ S \simeq 0.5 $.

Finally a quite simple model in reasonable agreement with known 
constraints is $f(y)=1/2-(1/ \pi ) \arctan ( \pi y/ 2 )$. Its linearization for 
$y \sim 0$ leads to take $ \xi = S = 0.5$. It gives the proper limit for 
the weakly interacting Fermi gas and in the BEC limit it yields $ a_m = 
6 a / \pi $ for the molecular scattering length, which is not too different 
from the result $ a_m = 0.6 a $ of Petrov \emph{et al} \cite{petrov}. 
Although it is very simple this model gives already a quite good 
agreement \cite{hu} with experiment \cite{grim} in the resonance
region as can  be seen in Fig.1. 

We are very grateful to F. Chevy, Y. Castin, C. Cohen-Tannoudji, J. 
Dalibard and C. Salomon for very stimulating discussions.

\noindent
* Laboratoire associ\'e au Centre National
de la Recherche Scientifique et aux Universit\'es Paris 6 et Paris 7.

\end{multicols}

\end{document}